%% file: itp_phot.tex
\documentclass[useAMS,usenatbib]{mn2e}
\usepackage{graphicx}

\newcommand{\Porb}{\mbox{$P_{\mathrm{orb}}$}}
\newcommand{\Line}[3]{\Ion{#1}{#2}\,$\lambda$\,#3}

\newcommand{\Ion}[2]{#1{\,\scriptsize #2}}

\newcommand{\id}{\mbox{$\mathrm{d^{-1}}$}}
\newcommand{\gsdss}{\mbox{$g_\mathrm{SDSS}$}}

\title[Photometric Observations of 15 SDSS CVs]{Orbital periods of
  cataclysmic variables identified by the SDSS. III. Time-series
  photometry obtained during the 2004/5 International
  Time Project on La Palma}
\author[M. Dillon, et al]{
M. Dillon$^1$,
B.T. G\"ansicke$^1$,
A. Aungwerojwit$^{2,1}$,
P. Rodr{\'i}guez-Gil$^3$,
T.R. Marsh$^1$,\newauthor
S.C.C. Barros$^1$,
P. Szkody$^4$,
S. Brady$^5$,
T. Krajci$^6$,
A. Oksanen$^7$\\
$^{1}$ Department of Physics, University of Warwick, Coventry CV4 7AL,
UK \\
$^2$ Department of Physics, Faculty of Science, Naresuan University, 
Phitsanulok, 65000, Thailand\\
$^3$ Instituto de Astrof{\'\i}sica de Canarias, 38200 La Laguna,
Tenerife, Spain\\
$^4$ Astronomy Department, University of Washington, Seattle, WA98195, USA\\
$^5$ 5 Melba Drive, Hudson, NH 03051, USA\\
$^6$ Astrokolkhoz Observatory,  1351 Cloudcroft, NM 88317, USA \\
$^7$ Hankasalmi Observatory, Kyllikinkatu 1, FI-40100 Jyv\"askyl\"a, Finland
}

\begin{document}

\date{Accepted 2008 Month day. Received 2008 Month day; in original form 2008 Month day}

\pagerange{\pageref{firstpage}--\pageref{lastpage}} \pubyear{2008}

\maketitle

\label{firstpage}

\begin{abstract}
We present time resolved CCD photometry of 15 cataclysmic variables
(CVs) identified by the Sloan Digital Sky Survey (SDSS). The data were
obtained as part of the 2004/05 International Time Programme on La
Palma. We discuss the morphology of the light curves and the CV
subtypes and give accurate orbital periods for 11 systems. Nine
systems are found below the 2--3\,h orbital period gap, of which five
have periods within a few minutes of the $\sim80$\,min minimum orbital
period. One system each is found within and above the gap. This confirms
the previously noted trend for a large fraction of short-period
systems among the SDSS CVs.  Objects of particular interest are
SDSS\,J0901+4809 and SDSS\,J1250+6655 which are deeply
eclipsing. SDSS\,J0854+3905 is a polar with an extremely modulated
light curve, which is likely due to a mixture of cyclotron beaming and
eclipses of the accretion region by the white dwarf. One out of five
systems with white-dwarf dominated optical spectra exhibits non-radial
pulsations.
\end{abstract}


\section{Introduction}
Cataclysmic variables (CVs) contain white dwarfs accreting from
(quasi) main-sequence stars. Because of their large number and the relatively
simple nature of their stellar components, CVs hold a great potential
to test and improve our understanding of compact binary evolution, in
particular of orbital angular momentum loss and the impact of mass
loss on the structure of the donor star. The severe disagreements
between the predictions of the ``standard'' theory of CV evolution
\citep[e.g.][]{rappaportetal83-1, kolb93-1, politano96-1,
howelletal01-1} and the properties of the observed CV population
\citep{patterson84-1, gaensicke05-1} have stimulated a substantial
number of alternative/additional theoretical approaches over the last
two decades \citep[e.g.][]{sharaetal86-1, clemensetal98-1,
kingetal02-1, schenkeretal02-1, willemsetal05-1}. However, at present
a quantitative test of these efforts is unfeasible, as the $\sim650$
moderately well studied CVs \citep{ritter+kolb03-1} represent an
inhomogeneous and incomplete population discovered by a number of
different methods \citep[e.g][]{gaensicke05-1,pretoriusetal07-1}.

The Sloan Digital Sky Survey (SDSS; \citealt{yorketal00-1}) has the
potential to dramatically improve the observational side of CV
population studies. Sampling a large volume in $ugriz$ colour space
and extending deeper than any previous large-scale survey, SDSS
provides the most homogeneous and complete sample of CVs to date.  At
the time of writing, the sample of definite SDSS CVs contains
$\sim220$ systems, of which $\sim180$ are new discoveries
(\citealt{szkodyetal02-2, szkodyetal03-2, szkodyetal04-1,
szkodyetal05-1, szkodyetal06-1, szkodyetal07-2} [henceforth PSI--PSVI]
\citealt{roelofsetal04-1, andersonetal05-1, schmidtetal07-1}, Groot et
al. 2007, submitted).

The newly identified CVs have been followed-up by a number of groups
\citep[e.g.][]{wolfeetal03-1, pretoriusetal04-1, woudt+warner04-1,
peters+thorstensen05-1, roelofsetal05-1, tramposchetal05-1,
gaensickeetal06-1, southworthetal06-1, southworthetal07-1,
southworthetal07-2, littlefairetal06-1, littlefairetal06-2}. Our group
has been awarded the International Time 2004/5 on La Palma for the study of
SDSS CVs, and we report in this paper photometric time series obtained
for 15 systems. Spectroscopic observations of 15
systems will be reported in a second paper (Dillon et al. in prep),
and a detailed discussion of the overall properties of the SDSS CV
sample will be given in a third paper (G\"ansicke et al. in prep).  In
Sect.\,\ref{s-observations}, we provide a brief description of the
instrumentation and data reduction used, and in Sect.\,\ref{s-tsa} we
outline the adopted time series analysis methods. Results on
individual objects are presented in Sect.\,\ref{s-results}, and a
general discussion of our results is given in
Sect.\,\ref{s-discussion}.

\input{obslog.tex}

\section{Observations}
\label{s-observations}
The bulk of our differential CCD photometry was obtained using five
different telescopes on La Palma as part of the 2004/5 International
Time Programme, with some additional data obtained at Calar Alto and
three small-aperture telescopes. We
selected as targets for photometric observations primarily systems
that were too faint for spectroscopy with the 2--4\,m telescopes
available on La Palma. A
total of 15 SDSS CVs were observed, 12 of which on more than one occasion
(Table\,\ref{t-obslog}). All data were reduced using the photometry
pipeline described by \citet{gaensickeetal04-1}. In brief, the images
were bias and flat-field corrected in \texttt{MIDAS}, aperture
photometry was performed using the \texttt{Sextractor}
\citep{bertin+arnouts96-1}, and differential magnitudes were measured
relative to a nearby comparison star. Differential magnitudes were
then converted to apparent magnitudes using the $g'$ magnitude of the
comparison star obtained from the SDSS data base. While data obtained
in the $V$-band can be calibrated using the colour transformations
available on the SDSS web site, some systematic uncertainty on the
apparent magnitudes of the systems observed in white light or with a
non-standard filter remains. Below we provide a brief description of
the instrumentation used at the different telescopes.

\textbf{Isaac Newton Telescope (INT).}  The Wide Field Camera (WFC) is
an optical mosaic camera used at the prime focus of the 2.5\,m INT. It
consists of four $2048\times4100$ pixel EEV CCDs with a
$0.33\,\arcsec\mathrm{pixel}^{-1}$ scale. Binning and windowing is not
supported and hence all four full CCD frames have to be read out,
which causes a dead-time of 42\,s between two exposures.

\textbf{William Herschel Telescope (WHT).}  The Auxiliary Port Imaging
Camera (AUX Port) is mounted at the Cassegrain auxiliary port of the
4.2\,m WHT. It is equipped with a $1024\times1024$\,pixel TEK CCD with
a $0.11\,\arcsec\mathrm{pixel}^{-1}$ scale and provides an unvignetted 
field of view of $\sim1.8\,\arcmin$. We used the AUX Port with a
binning $4\times4$, decreasing the read-out time of the full CCD to a
few seconds.  

\textbf{Liverpool Telescope (LT).} The LT is a fully robotic 2\,m
telescope. The optical CCD camera RATCAM is equipped with a
$2048\times2048$ pixel EEV  with $0.14\,\arcsec\mathrm{pixel}^{-1}$ 
scale covering a field of view of $4.6\,\arcmin\times4.6\,\arcmin$. A binning
$2\times2$ is the default for this instrument. Windowing is not
supported, and the read-out time for the full CCD is 10\,s. 

\textbf{Nordic Optical Telescope (NOT).}  The low resolution imaging
spectrograph ALFOSC on the 2.5\,m NOT telescope contains a
$2048\times2048$ pixel EEV CCD with an
$0.19\,\arcsec\mathrm{pixel}^{-1}$ scale. The data were obtained
binning the CCD $2\times2$ and windowing closely around the target
star, reducing the read-out time to a few seconds.

\textbf{Telescopio Nazionale Galileo (TNG).} The Device Optimized for the
Low Resolution (DOLORES) on the 3.6\,m TNG uses a $2048\times2048$ pixel
Loral CCD with a $0.28\,\arcsec\mathrm{pixel}^{-1}$ scale, covering a
field of view of about $9.4\,\arcmin\times9.4\,\arcmin$. The CCD was
not binned but windowed in order to reduce the readout time to
$\sim10$\,s.  

\textbf{Calar Alto (CA22).} The Calar Alto Faint Object Spectrograph (CAFOS)
was used on the 2.2\,m telescope. The instrument is equipped with a
$2048\times2048$ pixel SITe CCD with a scale of
$0.53\,\arcsec\mathrm{pixel}^{-1}$ and a field of view of
$16\,\arcmin$. Windowing was applied to reduce the readout time to
$\sim10$\,s. 

\textbf{Hankasalmi Observatory (HaO).} We used a RCOS Carbon 16RC\,0.40
m Ritchey-Chretien telescope mounted on a Paramount ME, along with an
SBIG~STL-1001E CCD camera. The data were dark subtracted and flat
fielded. Aperture photometry was then performed with the MaxImDL
software.

\textbf{Astrokolkhoz Observatory (AO).} We used a 0.28\,m
Schmidt-Cassegrain telescope with a focal length of 1.8\,m along with
an SBIG~ST-7 CCD camera binned $2\times2$. The data were dark and
bias subtracted and flat fielded and measured using aperture
photometry using AIP4WIN.

\textbf{Hudson Observatory (HO).}  Observations were made from a
private observatory located in southern New Hampshire,
USA. Instruments include a 0.4\,m robotic Newtonian Telescope mounted
on a Paramount ME along with an SBIG ST-8XME CCD camera, and BVRI
filters (which were not utilised due to the faint magnitude). The
robotic system was programmed to monitor a list of poorly studied CV
candidates and to initiate immediate time-series observations if an
outburst is detected, as was the case with SDSS\,J090103.93+480911.1
and SDSS\,J125023.84+665526.4. All images were bias subtracted,
flat-field corrected and photometrically reduced using MaxIm DL.

\begin{figure*}
\centerline{\includegraphics[width=178mm]{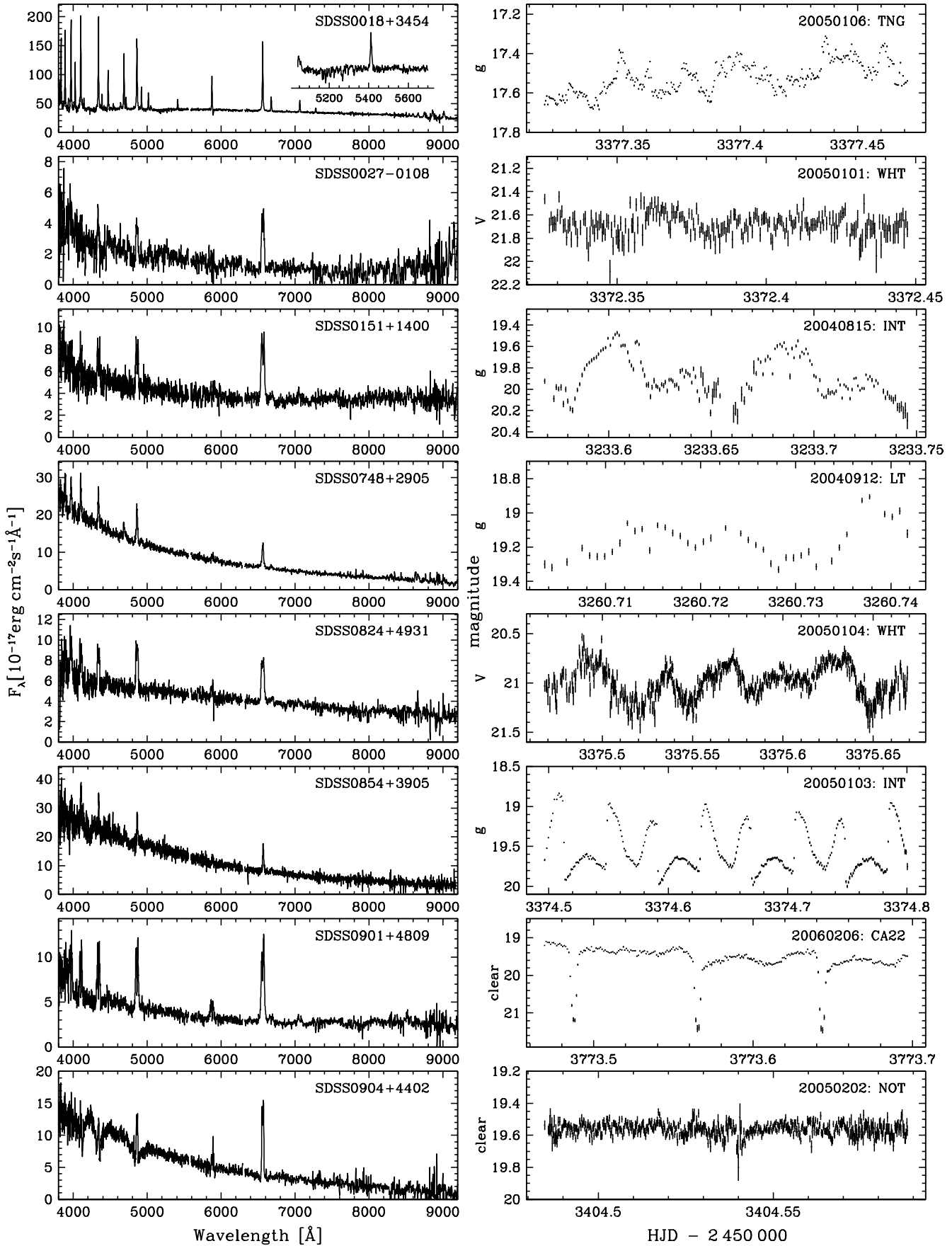}}
\caption{\label{f-lc}Sample light curves for all objects observed
(Table\,\ref{t-obslog}) along with their SDSS identification spectra
  from PSI--V.}
\end{figure*}

Sample light curves for each object are shown in Fig.\,\ref{f-lc},
where we also display the SDSS identification spectra from PSI--V for
convenience. At a first visual inspection, the light curves of the
observed systems display a wide variety of morphologies, including
apparently non-periodic flickering (SDSS\,J0018+3454), deep eclipses
(SDSS\,J1250+6655, SDSS\,J0901+4809), periodic double-humps
(e.g. SDSS\,J0151+1400), double-peaked flare-like events
(SDSS\,J0854+3905), or absence of significant variability altogether
(SDSS\,J1514+4549, SDSS\,J0904+4402).  The light curves obtained on
different nights for an individual system are rather similar in all
cases, except for SDSS\,J0901+4801, SDSS\,J1250+6655 and
SDSS\,J2116+1134, which were observed both in quiescence and in
outburst.

\begin{figure*}
\centerline{\includegraphics[width=178mm]{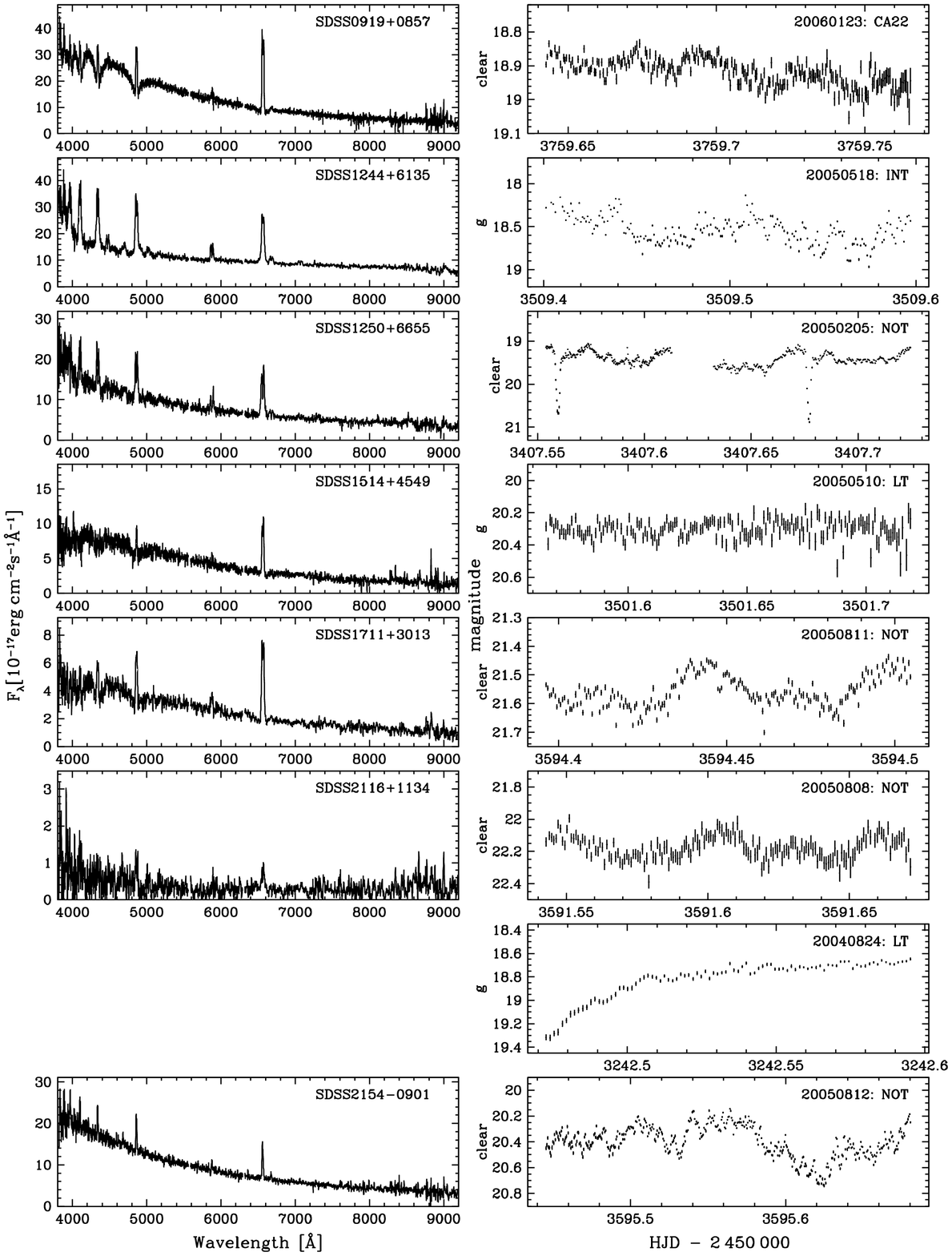}}
\textbf{Figure\,\ref{f-lc} continued.} Photometry of SDSS\,2116+1134  was
obtained both in quiescence and in outburst.
\end{figure*}

\section{Time Series Analysis}
\label{s-tsa}
All light curves were analysed using several methods provided by the
\texttt{MIDAS/TSA} context. More specifically, we computed
periodograms based on \citeauthor{scargle82-1}'s (1982) method
(\texttt{scargle}), as well as on
\citeauthor{schwarzenberg-czerny89-1}'s (1982) analysis-of-variance
(\texttt{aov}), and \citeauthor{schwarzenberg-czerny96-1}'s (1996)
extension of \texttt{aov} which fits the phase-folded data with
orthogonal trigonometric polynomials (\texttt{ort}). Fourier-type
methods such as \texttt{scargle} work best on light curves with
smooth, quasi-sinusoidal variations, but are not well-suited for the
analysis of light curves with sharp features, such as eclipses, where
\texttt{aov} and \texttt{ort} are preferred. A noticeable feature of
\texttt{aov} and \texttt{ort} periodograms is substantial power at 
sub-harmonics of the strongest signal.

We calculated periodograms using all three methods for each individual
light curve, as well as for each system combining all available
data. All light curves had their nightly mean subtracted before the
analysis. In 11 systems the time series analysis revealed the presence
of a periodic signal in the photometric data. We interpret these
signals as the orbital periods of the systems. The periodograms
calculated for those objects from their combined data sets are shown
in Fig.\,\ref{f-tsa}, along with phase-folded light curves. In order
to assess the likelihood that the strongest alias represents the true
orbital period we applied a test based on bootstrapping simulations
\citep[see][chapter 15.6]{pressetal92-1}, following the approach
described in \citet{southworthetal06-1, southworthetal07-2}. The
probabilities determined from this bootstrapping procedure are
slightly on the pessimistic side, as each simulation is using only a
subset of the entire available data. 

For the eclipsing systems SDSS\,J0901+4809 and SDSS\,J1250+6655 the
above methods were carried out for completeness. However, more
accurate orbital periods can be determined from the analysis of the
mid-eclipse times. In order to measure those times
(Table\,\ref{t-mideclipse}) we plotted the original light curve with a
copy mirrored in time, and shifted the copy until the bottom of the
eclipse in both light curves showed the closest agreement. Orbital
ephemerides were then determined from linear fits to the mid-eclipse
times. 

The periods and errors reported in Table\,\ref{t-obslog} were
determined from sine fits near the best-choice aliases in the
periodograms (Fig.\,\ref{f-tsa}), except for SDSS\,J0901+4809 and
SDSS\,J1250+6655, where the periods and errors were determined from
linear ephemeris fits to the mid-eclipse timings reported in
Table~\ref{t-mideclipse}.

\begin{table}
\caption[]{\label{t-mideclipse} Eclipse timings, cycle number, and the
difference in observed minus computed eclipse times using the
ephemerides in Eqs.~(\ref{eph-0901}) and (\ref{eph-1250}). }
\begin{tabular}{rrrr}
\hline
\multicolumn{1}{c}{Object} &
\multicolumn{1}{c}{$T_0$ (HJD)} & 
\multicolumn{1}{c}{$O-C$~(s)} &
\multicolumn{1}{c}{Cycle} \\
\hline
SDSS\,J0901+4801 &  2453773.48757 &  4 & 0 \\
                 &  2453773.56532 & -7 & 1 \\
                 &  2453773.64332 &  4 & 2 \\
                 &  2454379.78725 & -1 & 7785\\
                 &  2454379.86519 &  4 & 7786\\
                 &  2454382.90215 &-28 & 7825\\
                 &  2454382.98060 & 21 & 7826\\
                 &  2454383.36973 & -2 & 7831\\
                 &  2454383.44766 &  2 & 7832\\
                 &  2454384.84940 & -8 & 7850\\
                 &  2454384.92749 & 11 & 7851\\
\noalign{\smallskip}
SDSS\,J1250+6655 & 2453407.55964 &  -1 & 0   \\
                 & 2453407.67713 &   1 & 2   \\	
                 & 2453445.67918 &   5 & 649 \\
                 & 2453445.73774 & -10 & 650 \\
                 & 2453447.49997 &   4 & 680 \\
                 & 2453447.55857 &  -9 & 681 \\
                 & 2453447.61750 &   9 & 682 \\
                 & 2454494.58109 &   6 & 18507 \\
                 & 2454494.63971 &  -4 & 18508 \\
                 & 2454494.69853 &   3 & 18509 \\
                 & 2454494.75714 &  -8 & 18510 \\
                 & 2454494.81597 &   1 & 18511 \\
                 & 2454496.57856 &  46 & 18541 \\
                 & 2454496.63684 &   7 & 18542 \\
                 & 2454496.69558 &   7 & 18543 \\
                 & 2454496.75419 &  -4 & 18544 \\
                 & 2454496.81320 &  20 & 18545 \\
                 & 2454496.93046 &   1 & 18547 \\
                 & 2454500.68934 & -17 & 18611 \\
                 & 2454500.74828 &   2 & 18612 \\
\hline
\end{tabular}
\end{table}

\section{Results on individual systems}
\label{s-results}

In this Section, we present the results of our photometric
observations and discuss our findings for each individual system. 
A more generic discussion of these results follows in
Sect.\,\ref{s-discussion}.

\textbf{SDSS\,J0018+3454.} 
The SDSS identification spectrum in Fig.\,\ref{f-lc} displays strong
narrow emission lines. While the strength of \Line{He}{II}{4686} is
suggestive of a magnetic CV nature, the absence of noticeable
polarisation ($<0.2\%$) argues against SDSS\,J0018+3454 being a polar
(PSIV). Time-series spectroscopy with $3$\,\AA\ resolution obtained
over a 2\,h failed to detect a significant radial velocity variation
(PSIV). Our photometry obtained over four nights shows non-periodic
flickering activity with an amplitude of $\sim0.1$\,mag on time scales
of $\sim20-40$\,min and night-to-night variations of
$\sim0.3$\,mag. Close inspection of the SDSS spectrum of
SDSS\,J0018+3454 reveals a broad absorption dip centred on 5150\,\AA,
along with a host of narrow absorption lines in that range (see insert
in Fig.\,~\ref{f-lc}), typical of a mid-K main sequence star, strongly
suggesting an early-type donor star. In fact, the continuum spectrum
of SDSS\,J0018+3454 resembles closely that of
SDSS\,J204448.92--045928.8, for which \citet{peters+thorstensen05-1}
determined a donor star spectral type of K4--5 and an orbital period
of 2419\,min. The contribution of the donor star in SDSS\,J0018+3454
is weaker than in SDSS\,J2044--0459, indicating a relatively larger
contribution of the accretion luminosity in
SDSS\,J0018+3454. Combining the spectroscopic appearance with the
absence of noticeable radial velocity variations over short time
scales (PSIV), it appears likely that SDSS\,J0018+3454 is a
long-period CV, and phase-resolved spectroscopy spanning a
sufficiently long time should easily provide the radial velocity
variation of the donor star from its absorption lines. The 
strong \Line{He}{II}{4686} emission line in a presumably long-period
system is very untypical, and unravelling the nature of this object
should be high-priority task.

\begin{figure*}
\centerline{\includegraphics[width=170mm]{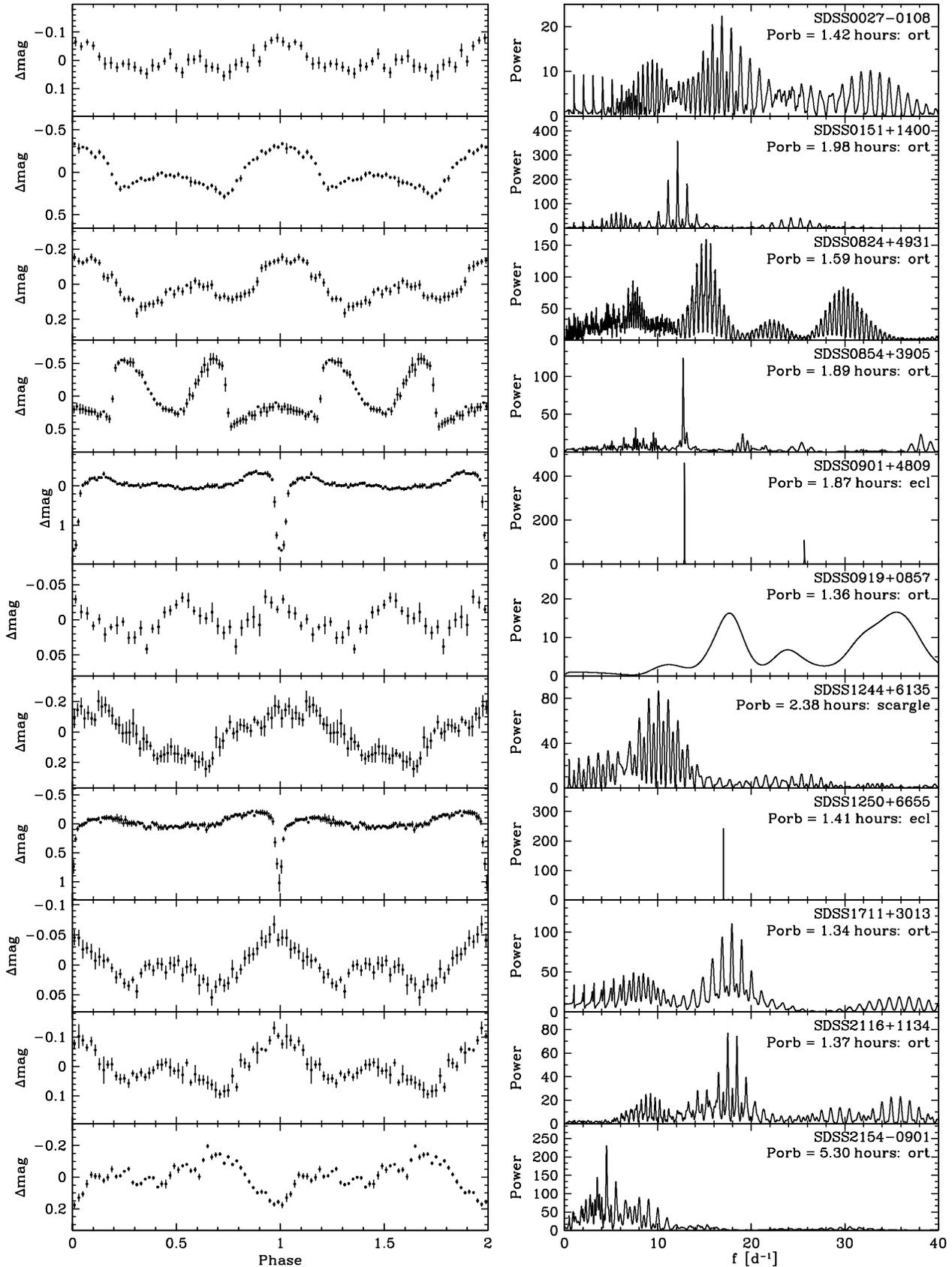}}
\caption{\label{f-tsa} Periodograms (right panels) and phase-folded
  light curves (left panels) of the 11 SDSS CVs for which we could
  determine an orbital period from our photometry
  (Table\,\ref{t-obslog}). The method indicated to compute the
  individual periodograms (\texttt{scargle}, \texttt{aov},
  \texttt{ort}) is indicated (see Sect.\,\ref{s-tsa}).}
\end{figure*}

\textbf{SDSSJ\,0027--0108 (EN\,Cet).} 
This object has been identified as a dwarf nova in outburst and
spectroscopically confirmed as a cataclysmic variable by
\citet{esamdinetal97-1}. The SDSS spectrum of EN\,Cet
(Fig.\,\ref{f-lc}, PSIV) is of rather poor quality, containing a
double-peaked H$\alpha$ emission line. The observed flux upturn at the
red end of the spectrum is interesting, as it could represent the
donor star in this object. However, it appears more likely that it is
an artifact from the night sky calibration, which is problematic in
fibre spectroscopy at such low flux levels. An \texttt{ort} analysis
of our $V$-band photometry obtained over three consecutive nights
reveals a periodic signal at 16.9\,\id\ (85.4\,min) and contains power
at the second harmonic as well. A bootstrap simulation
(Sect.\,\ref{s-tsa}) suggests a $\sim73$\% probability for the choice
of the strongest alias in the periodogram corresponding to the
intrinsic period of the data, the flanking 1\,\id\ aliases have
likelihoods of $\sim10$\%.  Phase-folding the data using this period
results in a light curve with a double-hump morphology
(Fig.\,\ref{f-tsa}).  Such a light curve shape is observed in a number
of ultrashort-period CVs, such as e.g. WZ\,Sge
\citep{pattersonetal98-2}, RZ\,Leo, BC\,UMa, MM\,Hya, \& HV\,Vir 
\citep{pattersonetal03-1}, WX\,Cet
\citep{mennickent94-1,rogoziecki+schwarzenberg-czerny01-1}, or
HS\,2331+3905 \citep{araujo-betancoretal05-1}), and we interpret
the 85.4\,min signal as the orbital period of the system. From a
sine-fit to the data, we determine $\Porb=85.44\pm0.07$\,min.

\textbf{SDSS\,J0151+1400.} 
The SDSS spectrum (Fig.\,\ref{f-lc}, PSI) contains double-peaked
Balmer emission lines, suggesting a moderately high inclination. The
He\,I lines are very weak, and the slope of the spectrum has a break
near 6500\,\AA, suggesting that the secondary star may contribute to
the observed flux in the red part of the SDSS spectrum. To explore
this possibility, we have applied a three-component model consisting
of a blackbody (to represent the blue continuum), an
isothermal/isobaric hydrogen slab (\citealt{gaensickeetal99-1}, to
represent the accretion disc), and an M-star template
(\citealt{rebassa-mansergasetal07-1}, to represent the secondary
star). A reasonable fit to the SDSS spectrum of SDSS\,J0151+1400 is
achieved for a blackbody temperature of 8500\,K, and a radius of the
blue component of $1.4\times10^9$\,cm, a slab temperature and column
density of 5800\,K and $2\times10^{-2}\,\mathrm{g\,cm^{-2}}$, and a
spectral type of the donor of M6, with all three components scaled to
a distance of 480\,pc (see \citealt{gaensickeetal06-1,
rodriguez-giletal05-1, southworthetal06-1} for more details on this
type of spectral modelling). These parameters should be considered
with some caution, as the SDSS spectrum is of rather poor quality and
covers an unknown orbital phase. Nevertheless, the spectral type
suggested by our fit is consistent with the donor stars that are
typically observed in this orbital period range
\citep[e.g.][]{silberetal04-1, remillardetal94-2,
mennickentetal02-2}. The geometric extension and temperature of the
blue continuum component, combined with the absence of broad Balmer
absorption lines, argues against an origin on the white dwarf but
rather for an origin from the accretion disc edge and/or bright
spot. A single outburst of SDSS\,J0151+1400 has been detected by
Hiroyuki Maehara in January 2007 (vsnet-alert 9139).

The $g$-band light curves obtained on three consecutive nights display
a double-humped morphology with an amplitude of $\sim0.3$\,mag. The
strongest signal in an \texttt{ort} periodogram is detected at a
frequency of 12.1\,\id\ (118.8\,min), with some power at the second
harmonic. A bootstrap simulation shows that the alias choice is
unambiguous for SDSS\,J0151+1400. In analogy to SDSS\,0027--0108, we
interpret the strongest signal as the orbital period of the system,
and obtain from a sine-fit to the data $\Porb=118.68\pm0.04$\,min.

\textbf{SDSS\,J0748+2905.}  
While the identification spectrum displays \Ion{He}{II} emission
(Fig.\,\ref{f-lc}, PSIII) the absence of noticeable polarisation ruled
out a polar nature for this object (PSIII). An orbital period of
2.5\,h was estimated by PSIII from radial velocity variations, though,
this must be considered a rough estimate only as the system was
observed only for $\simeq2.6$\,h. Our single $g$-band light curve
covering $\sim1$\,h shows variability on time scales of 15--20\,min
with an amplitude of $\simeq0.2$\,mag. We found the system at a mean
magnitude of $g\simeq19.2$, whereas the light curve in PSIII showed
the system at $V\simeq18.3$, indicating long-term variability of the
object (the nearby companion unresolved in the photometry reported in
PSIII has $g=22.5$ and can safely be neglected). No coherent periodicity was
detected in our data.

\textbf{SDSS\,J0824+4931.} 
The SDSS spectrum of SDSS\,J0824+4931 (Fig.\,\ref{f-lc}, PSI) broadly
resembles that of SDSS\,J0027--0108 and SDSS\,J0151+1400 with slightly
double-peaked Balmer emission lines along with very weak \Ion{He}{I}
lines. The secondary star does not noticeably contribute at red
wavelengths. Our $V$-band light curves are of a double-humped shape with
a $\sim0.5$\,mag amplitude. We observed SDSS\,J0824+4931 on three
consecutive nights (Table\,\ref{t-obslog}), with very poor conditions
in the second night. During our observations, SDSS\,J0824+4931 was
substantially fainter than the magnitude reported from the SDSS
imaging data.  Using only the data from the first and the third night,
an \texttt{ort} analysis results in an unambiguous signal at
15.16\,\id (\ref{f-tsa}), and a sine-fit to the data gives
$\Porb=94.99\pm0.02$\,min. In contrast, an \texttt{ort} analysis of
the combined data from all three nights (not shown) gives three possible
frequencies: 14.67\,\id\ (98.1\,min), 15.16\,\id\ (95.0\,min), and
15.65\,\id\ (92.0\,min). These three signals have roughly equal
strength in the power spectrum, and similar probabilities in a
bootstrap simulation. 

\citet{boydetal06-1} recently performed the first photometric study of
SDSS\,J0824+4931 during a superoutburst, detecting superhumps with a
period of 100.14$\pm$0.07\,min and an underlying weak signal at
98.9$\pm$0.9\,min, which they interpreted as the orbital period of the
system, suggesting a superhump excess of $\epsilon=0.13$. Using the
superhump period from \citet{boydetal06-1}, our three possible orbital
period aliases, 98.1\,min, 95.0\,min, and 92.0\,min, result in
$\epsilon=0.021$, 0.054, 0.088, respectively. Based on Patterson's
(\citeyear{pattersonetal05-3}) compilation of known superhump
excesses, the 98.1\,min alias seems most probably, as it implies a
low, but not unusual superhump excess, similar to e.g. KV\,And
\citep{pattersonetal03-1} and HS\,0417+7445
\citep{aungwerojwitetal06-1}. The 95.0\,min the 98.1\,min periods seem
less likely and implausible, as no dwarf novae with $\epsilon>0.05$
are found below the period gap \citep[e.g.][]{nogamietal01-1,
pattersonetal03-1, pattersonetal05-3}

In summary, as the exact value of the orbital period of
SDSS\,J0824+4931 remains somewhat debatable we prefer to err on the
conservative side, and report in Table\,\ref{t-obslog} an error
including the two flanking aliases from our photometry, hence
$\Porb=95\pm3$\,min.

\textbf{SDSS\,J0854+3905 (EUVE\,J0854+390).}
This system was identified as a possible magnetic CV by
\citet{christianetal01-1}. Their optical spectra displayed large
radial velocity variations suggesting a period of $<0.2$\,d. The SDSS
spectrum (Fig.\,\ref{f-lc}, PSIV) shows the system fainter by a factor
$\sim10$ with much weaker \Ion{He}{II} emission.  Spectropolarimetric
observations of SDSS\,J0854+3905 confirm the magnetic nature of the
system, displaying circular polarisation levels of up to $-30\%$ with
cyclotron harmonic humps clearly visible in the spectrum (PSIV).

Our two $g$-band light curves of SDSS\,J0854+3905 show a strong
periodic double-peaked flare-like morphology with a peak-to-peak
amplitude of $\sim1$\,mag with a period of $\sim2$\,h.  An
\texttt{ort} periodogram calculated from our data gives an unambiguous
signal at 12.7\,\id\ (113\,min), and a sine-fit (including four
harmonics because of the sharply modulated light curve morphology)
results in $\Porb=113.26\pm0.03$\,min. 

The extreme shape of the modulation can be understood in terms of a
combination of cyclotron beaming from the accretion column near the
white dwarf and the accretion region being eclipsed by the white dwarf
for parts of the spin cycle, similar to the geometry in AM\,Her
\citep{gaensickeetal01-2}. Maximum flux occurs near the orbital phases
$\varphi\simeq0.26$ and $\varphi\simeq0.68$ (Fig.\,\ref{f-tsa}, where
the zero-point of the phase is arbitrary), where the line-of-sight
must be closest to being perpendicular to the magnetic field lines in
the accretion column. The sharp rise and drop in flux observed near
phases $\phi\simeq0.2$ and $\phi\simeq0.75$ recurs with high precision
in both nights, and suggests that the accretion column/region is
self-eclipsed by the body of the white dwarf. The drop in flux centred
on $\phi\simeq0.5$ occurs when the magnetic pole rotates into view,
and the angle between the line-of-sight and the magnetic field lines in
the accretion region reaches a minimum value. The modulation observed
between $\phi\simeq0.75-0.20$ could be related to accretion onto the
second pole, or a geometric projection effect of the accretion
stream. Given that the white dwarf spin in polars is magnetically
locked to the orbital period, we deduce
$\Porb=113.26$\,min. SDSS\,J0854+3905 appears to be a promising object
for (spectro)polarimetric follow-up observations, which will allow a
precise reconstruction of the accretion geometry.

\textbf{SDSS\,J0901+4809.} 
The SDSS spectrum (Fig.\,\ref{f-lc}, PSII) shows deep central
absorption structures in the Balmer and \Ion{He}{I} emission lines,
suggesting a high inclination of the system. The slope of the spectrum
displays a break near 6500\,\AA, similar to SDSS\,J0151+1400. Applying
the same type of three-component model as for SDSS\,J0151+1400, we
find that rather similar parameters provide a satisfactory fit to the
SDSS spectrum of SDSS\,J0901+4809, i.e. a blackbody temperature and
radius of 9500\,K and $1.2\times10^9$\,cm, respectively, a slab
temperature and column density of 6100\,K and
$2\times10^{-2}\,\mathrm{g\,cm^{-2}}$, respectively, and a donor star spectral type
of M6, all scaled to a distance of 520\,pc. All in all,
SDSS\,J0151+1400 and SDSS\,J0901+4809 seem to be very similar in their
parameters with the latter one having a slightly higher inclination. 

Our single filterless light curve of SDSS\,J0901+4809
(Table\,\ref{t-obslog}) identifies the system as deeply eclipsing
one. On 6 Oct 2007, one of us (SB) detected the first outburst of
SDSS\,J0901+4809, finding the object at an unfiltered magnitude of
$\sim16.2$. Over the course of 5 days, it faded to an unfiltered
magnitude of $\sim18.1$, after which it was too faint for the
equipment available at that time. We measured the eclipse centres by
mirroring eclipse profiles and shifting them in time until the best
overlap was achieved. The eclipse timings are reported in
Table\,\ref{t-mideclipse}. A linear fit to the three eclipse times
results in the eclipse ephemeris

\begin{equation}                       
T_0(\mathrm{HJD})=2\,453\,773.48752(3)+0.077880505(35)\times E
\label{eph-0901}
\end{equation}

i.e. an orbital period of 112.15\,min $(12.84\,\id)$. The
corresponding cycle numbers and observed-minus-computed (O-C) eclipse
times are reported in Table\,\ref{t-mideclipse}. There is a small
chance for a cycle miscount, which would result in a somewhat shorter
period of 0.077870588(50)\,d (112.13\,min), however, the O-C values
prefer the  period given in Eq.\,\ref{eph-0901}.
SDSS\,J0901+4809 is a prime candidate for high time resolution eclipse
studies to measure accurate binary parameters \citep[see
e.g.][]{felineetal04-2, littlefairetal06-1}.

\textbf{SDSS\,J0904+4402.} 
The identification spectrum of SDSS\,J0904+4402 (Fig.\,\ref{f-lc},
PSIII) shows that the white dwarf is the dominant source of light in
this system. No trace of the donor star is detected in the red part of
the spectrum. Mildly double-peaked Balmer emission lines are
superimposed on the broad absorption lines from the white dwarf. Our
$g$-band and filterless photometry of SDSS\,J0904+4402 does not
contain any periodicity that could be identified as the orbital
period.

Inspired by the identification of a number of accreting white dwarfs
among the SDSS CVs which exhibit non-radial pulsations
\citep{woudt+warner04-1, warner+woudt04-1, gaensickeetal06-1,
nilssonetal06-1}, we have inspected the periodogram of
SDSS\,J0904+4402 for the presence of short-period signals. The INT
$g$-band data contain a (by means of a Monte-Carlo simulation)
significant signal at 8.678(34)\,min ($165.94(65)\id$) with an
amplitude of 26\,mmag. The NOT data contain two significant signals at
7.678(40)\,min ($187.53(99)\id$) and 12.16(11)\,min ($118.8(1.1)\id$),
with amplitudes of 12.5\,mmag and 11.3\,mmag, respectively. However,
given the fact that the two runs yield different periods, we would
consider SDSS\,J0904+4402 only as a pulsating WD candidate, as
accretion-related flickering could easily mimic such a putative
pulsation signal. Additional time-series photometry is encouraged to
test whether or not the periods detected at the INT and NOT are
detected again.

\textbf{SDSS\,J0919+5028.} 
The SDSS spectrum of SDSS\,J0919+5028 (Fig.\,\ref{f-lc}, PSIV) is
dominated by the white dwarf with no signature from the donor
star. PSIV estimated $\Porb\simeq1.4$\,h from 2\,h of time-resolved
spectroscopy, and compared the system to GW\,Lib, the first ZZ Ceti
star found in a CV \citep{vanzyletal00-1}. 

Our single filterless light curve shows humps with a period of
$\sim40$\,min superimposed on a downward trend in the system
brightness. We detrended the overall trend by subtracting from the
light curve a copy of the data smoothed with an 80-point box car.  An
\texttt{ort} periodogram computed from the detrended light curve
contains broad peaks centred at a frequency of 17.7\,\id\ and
its harmonic, suggesting that the actual morphology of the light curve
is double-humped, similar to SDSS\,J0027--0108, SDSS\,J0151+1400, and
SDSS\,J0824+4931. A sine-fit to the data results in
$\Porb=81.6\pm1.2$\,min.


The power spectrum at shorter periods reveals a marginally significant
signal at $259.93\pm0.94$\,s ($332.41\id$), which we interpreted as a
possible non-radial pulsation mode.  Based on the detection of that
signal in additional data, \citet{mukadametal07-1} confirmed
SDSS\,J0919+5028 as a white dwarf pulsator.

\textbf{SDSS\,J1244+6135.} 
The SDSS spectrum (Fig.\,\ref{f-lc}, PSII) displays strong
double-peaked Balmer and \Ion{He}{I} emission lines, suggesting a high
inclination. Our $g$-band data obtained over three consecutive nights
reveal a periodic modulation with an amplitude of $\simeq0.2$\,mag. A
\texttt{scargle} periodogram contains a cluster of aliases with the
strongest signal at 10.0\,\id\ (144\,min). The absence of power at the
second harmonic indicates that the light curve morphology of
SDSS\,J1244+6135 is not, as in e.g. SDSS\,J0027--0108 or
SDSS\,J0151+1400, double-humped. A bootstrap simulation shows that the
alias choice is unambiguous with a likelihood of 97.2\%. We interpret
the observed modulation as the orbital period, and determine
$\Porb=142.9\pm0.2$\,min.

\textbf{SDSS\,J1250+6655.} 
The identification spectrum of SDSS\,J1250+6655 (Fig.\,\ref{f-lc},
PSII) contains broad double-peaked Balmer emission lines with deep
central absorption, typical of high-inclination systems. No
spectroscopic feature from the donor star is detected in the red part
of the spectrum.  PSII suggested an orbital period of 5.6\,h on the
basis of a short (2.3\,h) set of time-resolved spectroscopy, but the
long (15 min) integration times could not resolve eclipses.  Our NOT
light curve of SDSS\,J1250+6655 revealed the occurrence of deep
($\sim2$\,mag) eclipses. A first outburst of this CV was detected by
one of us (SB) on 29 January 2008. The mid-eclipse times
(Table\,\ref{t-mideclipse}) were determined from all our light curves
in an analogous way as for SDSS\,J0901+4809, and a linear fit to those
times gives the following eclipse ephemeris

\begin{equation}
T_0(\mathrm{HJD})=24\,534\,07.55966(7)+0.0587356870(43)\times E, 
\label{eph-1250}
\end{equation}

i.e. an orbital period of 84.5793893(63)\,min ($\simeq17.0\id$).  With
SDSS\,J1250+6655 being close to the orbital minimum period, eclipse
studies appear especially worthwhile to determine accurate masses and
radii for the stellar components in this system~--~lying between the
eclipsing XZ\,Eri ($\Porb=88.0$\,min, $M_2=0.0842\pm0.0024\,M_\odot$,
\citealt{felineetal04-2}) and SDSS\,J1035+0551 ($\Porb=82.1$\,min,
$M_2=0.052\pm0.002\,M_\odot$, \citealt{littlefairetal06-2,
southworthetal06-1}).

\textbf{SDSS\,J1514+4549.} 
This is another white dwarf dominated CV, with no noticeable
contribution from the donor star (Fig.\,\ref{f-lc}, PSIV). The two
$g$-band light curves reveal very little variability, and no
periodicity that could be ascribed to the orbital period is detected
in the periodograms calculated from the data. Our two observations
found the object at a constant mean brightness of
$g\simeq20.2$. \citet{nilssonetal06-1} reported the detection of
non-radial pulsations with a period of 559.3\,s in two observations,
and the absence of this signal in a third observation. The
non-detection of this period in our two observing runs suggests that
SDSS\,J1514+4559, if it is a pulsator, has a rather unstable pulsation
spectrum.

\textbf{SDSS\,J1711+3013.} 
Very similar to SDSS\,J1514+4549, the SDSS spectrum of
SDSS\,J1711+3013 clearly reveals the white dwarf (Fig.\,\ref{f-lc},
PSIII) but no evidence for the secondary star. We obtained a total of
5 light curves on three consecutive nights in August 2004 and on two
consecutive nights in August 2005. All light curves reveal a
double-humped morphology. An \texttt{ort} periodogram of the 2005
data, which is of better quality than the 2004 data, contains the
strongest peak at 17.9\,\id\ (80.4\,min), which we interpret as the
orbital period. A bootstrap simulation shows that the choice of this
alias is unambiguous (99.6\% likelihood). A sine-fit to the data
results in $\Porb=80.35\pm0.05$\,min.  Combining the data from both
years confirms this result, but does not allow a more accurate period
determination as the periodogram suffers from strong 1-year aliases. We
note that no significant signal at shorter periods is found in any of
our observations, with a detection limit of 12\,mmag, which is
consistent with the non-detection of ZZ\,Ceti pulsations in this
object by \citet{mukadametal07-1}

\textbf{SDSS\,J2116+1134.} 
The SDSS spectrum of SDSS\,J2116+1134 (Fig.\,\ref{f-lc}, PSIII) shows
double-peaked Balmer emission lines on a very weak continuum. The
object was found in the SDSS imaging data at $g=15.3$, whereas we
measured $g=21.8$ from the SDSS spectrum, clearly showing that
SDSS\,J2116+1134 is a dwarf nova. In August 2004, we obtained
photometry with the LT and found the object rising in magnitude from
$g=19.4$ to 18.6 over the course of a 3\,h-long observing run,
indicating another outburst of the system. In August 2005, filterless
photometry obtained over three consecutive nights showed the system at
a mean magnitude of $g\simeq22.2$, displaying a double-humped light
curve. An \texttt{ort} periodogram contains two strong aliases at
17.5\,\id\ (82.3\,min) and 18.5\,\id\ (77.8\,min), with the bootstrap
test giving a 2/3 preference for the lower frequency alias. Sine fits
to both aliases determine the two periods to $\pm0.03$\,min, however,
given the ambiguous choice of the correct alias, we adopt
$\Porb=80.2\pm2.2$\,min as a conservative value.

\textbf{SDSS\,J2154--0901.}
The identification spectrum of SDSS\,J2154--0901 (Fig.\,\ref{f-lc},
PSII) shows a blue continuum superimposed by relatively weak and
narrow Balmer and \Ion{He}{I} lines, as well as weak
\Line{He}{II}{4686}. 
Our light curves of SDSS\,J2154--0901 obtained over three consecutive
nights show substantial flickering activity and a long term modulation
with a broad minimum and a steep decline. An \texttt{ort} periodogram
contains a strong signal at 4.5\,\id\ (320\,min), and a sine fit to
the data gives $\Porb=319\pm0.7$\,min. SDSS\,J2154--0901 is the only
system above the $2-3$\,h period gap among the 11 systems for which we
were able to determine periods.

\section{Discussion}
\label{s-discussion}
Substantial effort has gone into calculating models of the
intrinsic population of galactic CVs \citep[e.g.][]{dekool92-1,
kolb93-1, politano96-1, howelletal01-1}, however, comparison with the
observed properties of the known CV sample consistently failed to
match the predicted orbital period distribution, the total space
density of CVs, and the expected tight correlation between orbital
period and mass transfer rate \citep[e.g.][]{patterson84-1,
ringwald93-2, gaensicke05-1}. While these discrepancies may be
partially related to shortcomings in the theories of common envelope
evolution and orbital angular momentum loss, it is clear that the
heterogeneous set of known CVs \citep{ritter+kolb03-1} is not
well-suited for a quantitative test of the population models. 

SDSS is currently establishing the largest, deepest ($g\sim20$), and
most homogeneous sample of CVs, with an estimated $\sim300$ CVs by the
end of the SDSS operations. Establishing the detailed properties of
this CV sample is a major task.  Our photometric study of 15 CVs
contained within the SDSS spectroscopic data base has led to the
determination of the orbital periods for 11 of them, along with additional
information on CV subtypes. This underlines the potential of
time-series photometry on mid-size aperture telescopes for measuring
the orbital periods of the faint SDSS CVs.

Inspecting the distribution of orbital periods among the 11 CVs
studied here (Table\,\ref{t-obslog}), it becomes apparent that nine
systems have periods below the 2--3\,h period gap, one within the gap,
and only one above the gap. We were unable to determine the orbital
periods for four systems, but based on their spectroscopic appearance
the white-dwarf dominated SDSS\,J0904+4402 and SDSS\,J1514+4549 are
likely to have periods below the gap, and SDSS\,J0748+2905 and
SDSS\,J0018+3454 are likely to have periods above the gap.  Our result
corroborates the findings of \citet{szkodyetal03-2, szkodyetal07-2}
and \citet{southworthetal06-1, southworthetal07-2}, who noticed a
larger fraction of short-period systems among the SDSS CVs compared to
the previously known CVs (see e.g. \citealt{aungwerojwitetal06-1} for
CVs from the Hamburg Quasar Survey or \citealt{pretoriusetal07-1} for
CVs from the Palomar Green Survey).  In particular, five systems have
periods within $\sim5$\,min of the $\sim80$\,min orbital
minimum. Three of those have optical spectra dominated by the white
dwarf (SDSS\,J0027--0108, SDSS\,J0919+0857, and SDSS\,J1711+3013),
with no signature from the mass donor, indicating both very low mass
transfer rates and a very late spectral type of the companion
star. All three closely resemble SDSS\,J1035+0551, an eclipsing
white-dwarf dominated CV \citep{southworthetal06-1} with an 
low-mass brown dwarf donor \citep{littlefairetal06-2}. About
$\sim20$\% of the CVs identified by SDSS have white-dwarf dominated
spectra (PSI--VI) similar to SDSS\,J1035+0551 and the systems studied
here, and orbital period measurements are available for $\sim20$ of
these systems, with the majority being found near the minimum period
\citep[e.g.][]{woudt+warner04-1, pretoriusetal04-1, zharikovetal06-1,
gaensickeetal06-1, southworthetal06-1, southworthetal07-2}. Hence, it
appears likely that complete follow-up of the SDSS CV sample will lead
to a substantial increase of CVs with extremely short orbital periods.

A final note concerns the detection of non-radial pulsations in a number
of SDSS CVs \citep{woudt+warner04-1, gaensickeetal06-1,
nilssonetal06-1, mukadametal07-1}. While all of the confirmed
pulsators have white-dwarf dominated spectra, it does not appear possible
to predict the presence of pulsations in a given system just on the
base of its optical spectrum. Among the CVs studied here, five exhibit
the white dwarf in the SDSS spectrum, but only one system appears to
be a non-radial pulsator (SDSS\,J0919+0857). Ultraviolet observations
show that pulsations can appear over a wide range of white dwarf
effective temperatures \citep{szkodyetal02-4, araujo-betancoretal05-1,
szkodyetal07-1}, and hence the ``instability strip'' is apparently
less well-defined than for single white dwarfs \citep{mukadametal04-2,
gianninasetal06-1}. This difference may be related to the
contamination of the envelope by accreted helium in the CV white
dwarfs \citep{arrasetal06-1}.

\section{Conclusions}
We have obtained time-series photometry for 15 CVs contained in the
SDSS spectroscopic data base. Periodic variability detected in 11 of
these systems allowed us to establish their orbital periods.  We find
that 9 systems are located below the $2-3$\,h orbital period gap, one
within the gap, and one above, confirming that the SDSS CV sample
contains a larger fraction of short-period systems compared to the
previously known CVs.  Two of the new CVs are eclipsing,
SDSS\,J0901+4809 and SDSS\,J1250+6655, and are prime targets for
detailed binary parameter studies. One polar, SDSS\,J0854+3905,
exhibits a sharply modulated light curve and warrants polarimetric
follow-up in order to establish the accretion geometry. Finally, only
one out of five CVs having white-dwarf dominated spectra displays
non-radial pulsations. 

\section*{Acknowledgements}
MD was supported by an STFC Studentship. 
AA thanks the Royal Thai government for generous funding.
PS acknowledges some support from NSF grant AST 02-05875.
SCCB is supported by FCT.
Based in part 
on observations made with the William Herschel Telescope and the Isaac
Newton Telescope, which are operated on the island of La Palma by the
Isaac Newton Group in the Spanish Observatorio del Roque de los
Muchachos (ORM) of the Instituto de Astrof{\'\i}sica de Canarias (IAC);
on observations made with the Telescopio Nazionale Galileo 
operated on the island of La Palma by the Centro Galileo Galilei of
the INAF (Istituto Nazionale di Astrofisica) at the ORM of the IAC;
on observations made with the Nordic Optical Telescope, operated
on the island of La Palma jointly by Denmark, Finland, Iceland,
Norway, and Sweden, in the ORM of the IAC;
The Liverpool Telescope is operated on the island of La Palma by
Liverpool John Moores University at the ORM of the IAC.
The WHT, INT, NOT, TNG, and LT data were obtained as part of the 2004
International Time Programme of the night-time telescopes at the
European Northern Observatory.


\clearpage
\bsp
\label{lastpage}

\end{document}

%% file: obslog.tex
\begin{table*}
 \centering
\caption{\label{t-obslog} Log of observations and summary of the
system properties. We list the orbital periods as determined from our
photometric time series, the $g$ magnitude from the SDSS imaging data,
the magnitude during our observations, information on the system type
(WD\,=\,white-dwarf dominated spectrum, RD\,=\,secondary star visible,
\Ion{He}{II}\,=\,noticeable \Ion{He}{II} emission line,
EC\,=\,eclipsing, DN\,=\,dwarf nova, AM\,=\,polar), the date of the
observation, the telescope and filter used, the exposure time, the
number of images, and a reference to the identification paper.}
\setlength{\tabcolsep}{1.1ex}
\begin{tabular}{llrllcccccc}

  \hline
SDSSJ    & \Porb\,[min] &  \gsdss & mag & Type  & Date/Time [UT] & Telescope & Filter & Exp(s). &
   Frames & Source \\
 \hline
001856.93+345444.3 & - & 17.8 & 17.2 & \Ion{He}{II}/RD & 2005-01-01 20:55-23:46 & INT & clear & 25 & 153 &PSIV \\ 
                      &&      & 17.8 &    & 2005-01-05 19:50-23:01 & TNG & clear & 40 & 176 &   \\ 
                      &&      & 17.5 &    & 2005-01-06 19:35-23:17 & TNG & clear & 30 & 321 &   \\ 
                      &&      & 17.7 &    & 2005-01-07 19:29-22:48 & TNG & clear & 30 & 275 &   \\ 
002728.01-010828.5 & $85.44\pm0.07$ & 20.7 & 20.6 & DN/WD & 2004-12-31 20:08-22:55 & WHT & $V$ & 40-100 & 215 &PSIV\\ 
                           &&              & 20.6 &       & 2005-01-01 19:49-22:44 & WHT & $V$ & 40     & 233 &  \\ 
                           &&              & 20.6 &       & 2005-01-02 19:35-22:29 & WHT & $V$ & 40     & 231 &  \\ 
015151.87+140047.2 & $118.68\pm0.04$ &20.3& 19.9 & DN/RD & 2004-08-15 02:47-05:39 & INT & $g$ & 40    & 129 &PSI \\ 
                           &&             & 19.9 &  & 2004-08-16 01:35-05:50 & INT & $g$ & 40-60 & 169 &  \\ 
                           &&             & 19.9 &  & 2004-08-17 01:58-05:30 & INT & $g$ & 40-60 & 148 &  \\ 
074813.54+290509.2 & - & 18.6 & 19.2 & \Ion{He}{II} & 2004-09-12 04:57-05:53 &  LT & $g$ & 60 & 50 &PSIII  \\ 
082409.73+493124.4 & $95\pm3$ & 19.3 & 21.0 & DN & 2005-01-04 23:06-03:56 & WHT & $V$ & 20    & 698 & PSI \\ 
                             &&      & 20.8 &    & 2005-01-05 21:00-04:56 & WHT & $V$ & 30    & 380 &  \\ 
                             &&      & 21.0 &    & 2005-01-07 00:37-04:24 & WHT & $V$ & 20-30 & 532 &  \\ 
085414.02+390537.3 & $113.26\pm0.03$ &19.2 & 19.2 & AM & 2005-01-03 00:07-07:13 & INT & $g$ & 80-120 & 112 &PSIV\\ 
                                     &&    & 19.5 &    & 2005-01-03 23:47-07:13 & INT & $g$ & 50-60  & 303 &  \\ 
090103.93+480911.1 & 112.14793 & 19.9& 19.5 & DN/RD/EC & 2006-02-06 23:08-04:27 & CA22 & clear & 60  & 240 & PSIV \\ 
                & $\pm0.00005$ &     & 16.2 &          & 2007-10-06 06:52-09:39 & HO  & clear & 160 &  52 &      \\ 
                   &       &     & 17.0 &          & 2007-10-09 08:38-12:12 & AO   & clear &  60 & 192 &      \\ 
                   &       &     & 17.1 &          & 2007-10-09 20:42-00:22 & HaO  & clear &  95 & 128 &      \\ 
                   &       &     & 18.1 &          & 2007-10-11 08:10-12:09 & AO  & clear &  90 & 138 &      \\ 
090452.09+440255.4 & - &19.4& 19.6 & WD& 2005-01-05 02:33-04:06 & INT & $g$     & 30-40 & 73  &PSIII \\ 
                        &&  & 19.1 &   & 2005-01-05 04:10-06:55 & INT & clear & 30-35 & 143 &    \\ 
                        &&  & 19.6 &   & 2005-02-02 23:30-02:00 & NOT & clear &    15 & 544 &    \\ 
091945.11+085710.0 & $81.6\pm1.2$ &19.9 & 18.9 & WD & 2006-01-24 03:17-06:14 & CA22 & clear & 25 & 252 & PSIV\\ 
124426.26+613514.6 & $142.9\pm0.2$&18.8 & 18.6 & \Ion{He}{II} & 2005-05-18 21:37-02:20 & INT & $g$ & 40-60 & 215 &PSIII \\ 
                                   &&   & 18.4 &   & 2005-05-19 21:03-00:29 & INT & $g$ &    40 & 160 &    \\ 
                                   &&   & 18.7 &   & 2005-05-20 21:02-00:05 & INT & $g$ & 40-60 & 123 &    \\ 
125023.85+665525.5 & 84.5793893 &18.7 & 19.1 & DN/EC & 2005-02-03 04:19-05:38 & NOT & clear & 15-30 & 151 &PSII \\ 
                  & $\pm0.0000036$&  & 19.3 &    & 2005-02-05 01:14-05:19 & NOT & clear & 10-20 & 451 &   \\ 
                       &&          & 19.0 &    & 2005-03-16 03:15-06:42 & LT  & $g$     & 60    & 169 &   \\ 
                       &&          & 19.1 &    & 2005-03-17 23:19-23:47 & LT  & $g$     & 60    & 33  &   \\ 
                       &&          & 19.0 &    & 2005-03-18 01:13-03:14 & LT  & $g$     & 60    & 166 &   \\ 
		       &&          & 16.0 &    & 2008-01-29 01:36-08:52 & HO  & clear   & 70    & 306 &   \\ 
		       &&          & 16.4 &    & 2008-01-31 01:18-10:40 & HO  & clear   & 120   & 231 &   \\ 
		       &&          & 16.6 &    & 2008-02-04 03:32-08:06 & HO  & clear   & 160   &  93 &   \\ 

151413.72+454911.9 &- &19.7 & 20.1 & WD & 2005-04-01 04:11-06:29 & LT & $g$ & 60 & 114  &PSIV \\ 
                         && & 20.1 &    & 2005-05-10 01:30-05:35 & LT & $g$ & 60 & 205  &   \\ 
171145.08+301320.0 & $80.35\pm0.05$ &20.3 & 20.2 & WD & 2004-08-14 21:15-01:03 & INT & $g$ & 40-60  & 141 & PSIII \\ 
                               &&         & 20.2 &    & 2004-08-15 21:05-00:13 & INT & $g$ & 40-60  & 129 &     \\ 
                               &&         & 20.2 &    & 2004-08-16 21:05-23:46 & INT & $g$ & 45-120 &  63 &     \\ 
                               &&         & 20.3 &    & 2005-08-11 21:28-00:04 & NOT & clear & 40 & 198 &     \\ 
                               &&         & 20.3 &    & 2005-08-12 21:11-22:30 & NOT & clear & 40 & 101 &     \\ 
211605.43+113407.5 & $80.2\pm2.2$ &15.3& 22.1 & DN & 2005-08-09 00:54-03:59 & NOT & clear & 60 & 164 &PSIII \\ 
                             &&        & 22.1 &    & 2005-08-10 00:40-02:24 & NOT & clear & 60 &  93 &    \\ 
                             &&        & 22.1 &    & 2005-08-11 02:40-04:33 & NOT & clear & 90 &  70 &    \\ 
                             &&        & 18.8 &    & 2005-08-24 23:12-02:08 & LT & $g$ & 90 & 110  \\
215411.12-090121.6 & $319\pm0.7$& 19.2 & 20.4 && 2005-08-11 00:29-05:42 & NOT & $W$ & 40 & 396 & PSII\\ 
                      &&         & 20.3 && 2005-08-12 22:33-04:11 & NOT & $W$ & 40 & 427 &   \\ 
                      &&         & 20.4 && 2005-08-14 00:33-05:38 & NOT & $W$ & 40 & 376 &   \\ 
\hline 
\multicolumn{11}{c}{clear\,=\,white light, $V$\,=\,Johnson-$V$ band,
$g$\,=\,Sloan $g$-band, $W$\,=\,broad-band red cut-off.}
\end{tabular}

\end{table*}